\documentclass{jetpl}
\twocolumn

\rus

\usepackage{caption}
\usepackage{graphicx}
\usepackage{subcaption}
\usepackage{cite}
\usepackage[dvips]{graphicx}
\usepackage{multirow}
\DeclareGraphicsExtensions{.pdf,.png,.jpg,.eps,.ps}
\RequirePackage{caption}
\DeclareCaptionLabelSeparator{point}{. }
\makeatletter
\newcommand{\keepwithnext}{\@beginparpenalty 10000}
\makeatother

\title{Management of high-tech companies in conditions of import substitution}
\rtitle{Management of high-tech companies in conditions of import substitution}
\sodtitle{Management of high-tech companies in conditions of import substitution}

\author{S.\,E.\,Pyatovsky \/\thanks{~E-mail: vgsep@ya.ru}, N.\,S.\,Efimova, E.\,V.\,Surkova}
\rauthor{S.\,E.\,Pyatovsky}
\sodauthor{Pyatovsky}

\abstract{The article analyzes the development of high-tech sectors of the Russian economy in the context of import substitution. Features of managing priority project portfolios are considered. Issues of creating a unified information space for aviation industry enterprises are studied in the context of introduction of a modified OLAP technology of management decision support. Investment attractiveness of high-tech sectors of the Russian economy is estimated based on the coefficient of gross value added of project products. Investment-overheated industries are identified, and recommendations on market correction and returning project assets to a balanced state are given.\newline
}


\begin{document}

\maketitle
{\bf 1.~Introduction}\newline
The process of recovery of the Russian and global economies is as of 2024 still incomplete. The global economy is headed towards a prolonged recession with
signs of deindustrialization. Widespread introduction of high technologies has led to the fact that very few labor resources are willing to choose physical labor over intellectual. The latter has become one of the incentives for further development of high-tech companies (HTCs). The areas that can be considered major development growth points for high-tech industries are renewable energy and automotive and electric vehicles production.

HTCs determine the competitive ability of a nation’s economy. The production of HTCs is expected to ensure the level of research and development
(R\&D) required for the security of the national economy and independence of production from foreign suppliers, which is usually considered as an aspect
of import substitution.

Design companies are classified as high-tech via assessment of R\&D intensity of the production chain as the ratio of investments in R\&D to the gross value added of the company’s products. The management of the design company creates strategic project portfolios that are implemented at fundamental companies, for example, enterprises of the aviation industry.

{\bf 2.~Managing priority project portfolios}\newline
The composition of priority portfolios of import substitution projects is determined by the products of the following critical industries of the Russian economy. It is advisable to implement these project portfolios using the reverse engineering method, which involves design and development work based on finished product samples~\cite{1,2,3}.

For HTC management in the context of import substitution, in terms of research and development it is advisable to focus on projects with a small planning
horizon (so-called "quick projects"), which provide project results in the form of a replacement range of products stated in the project’s feasibility study within a period of up to one year. In terms of financing priority project portfolios, R\&D should be carried out under special investment contracts~\cite{4} with information and analytical support for investment projects~\cite{5} within the framework of a computer-aided design/automated economic information system (CAD-AEIS) for HTCs.

{\bf 3.~CAD-AEIS framework for aviation industry HTCS}\newline
Issues of creating a unified information space for aviation industry HTCs have been studied in the authors’ previous works~\cite{6,7}. To achieve maximum efficiency, the AEIS needs highly effective data analysis tools. Such tools include, for example, OLAP technologies~\cite{8,9} to be implemented at the HTC
(depending on the type of automation project, in M-, R-, or H-OLAP modification). Open source applications have made it possible for aviation industry HTCs
to use AEIS without having to invest in expensive licenses.

{\bf 4.~Investment attractiveness of high-tech sectors of the Russian economy}\newline
Production accounts by sectors of the economy~\cite{10,11}, analyzed based on statistical data, are shown in Table~\ref{tabl:1}, which is compiled with consideration to coefficient $K_\textrm{GVA}$ (equation~\ref{eq1}). The coefficient of gross value added of a project product is defined as:

\begin{equation}
K_\textrm{GVA}=GVA/OBP
\label{eq1}
\end{equation}

where $GVA$ is the gross value added~\cite{10,11} of the project product and $OBP$ is the output at basic prices~\cite{10,11} of the project product.

\setcounter{table}{0}
\renewcommand\thetable{\arabic{table}}
\begin{table*}[htbp]
\caption{}
  \centering
    \begin{tabular}[t]{|c|l|c|c|}
\hline    
\textit{N}&\multicolumn{1}{c|}{Industry categories according to~\cite{11}}&\textit{GVA}, bn RUB&\textit{OBP}, bn RUB\\
\hline
1&Education&3724&4712\\
\hline
2&Real estate activities&11711&14831\\
\hline
3&Mining and quarrying&15031&23265\\
\hline
4&Financial and insurance activities&5384&7621\\
\hline
5&Public administration and defence; compulsory social security&8404&12343\\
\hline
6&Administrative and support service activities&2214&3323\\
\hline
7&Human health and social work activities&3958&6051\\
\hline
8&Arts, entertainment and recreation&1066&1727\\
\hline
9&Professional, scientific and technical activities&5256&9034\\
\hline
10&Wholesale and retail trade; repair of motor vehicles and motorcycles&15270&26688\\
\hline
11&Other services&627&1369\\
\hline
12&Agriculture, forestry and fishing&4974&9603\\
\hline
13&Information and communication&3235&6314\\
\hline
14&Accommodation and food service activities&955&2097\\
\hline
15&Transportation and storage&7070&16176\\
\hline
16&Construction&5964&14529\\
\hline
17&Water supply; sewerage, waste management and remediation activities&657&1999\\
\hline
18&Electricity, gas, steam and air conditioning supply&2866&9704\\
\hline
19&Manufacturing&18926&68530\\
\hline
    \end{tabular}
\label{tabl:1}    
\end{table*}

\begin{figure}[h]
\centering
\includegraphics[width=\linewidth]{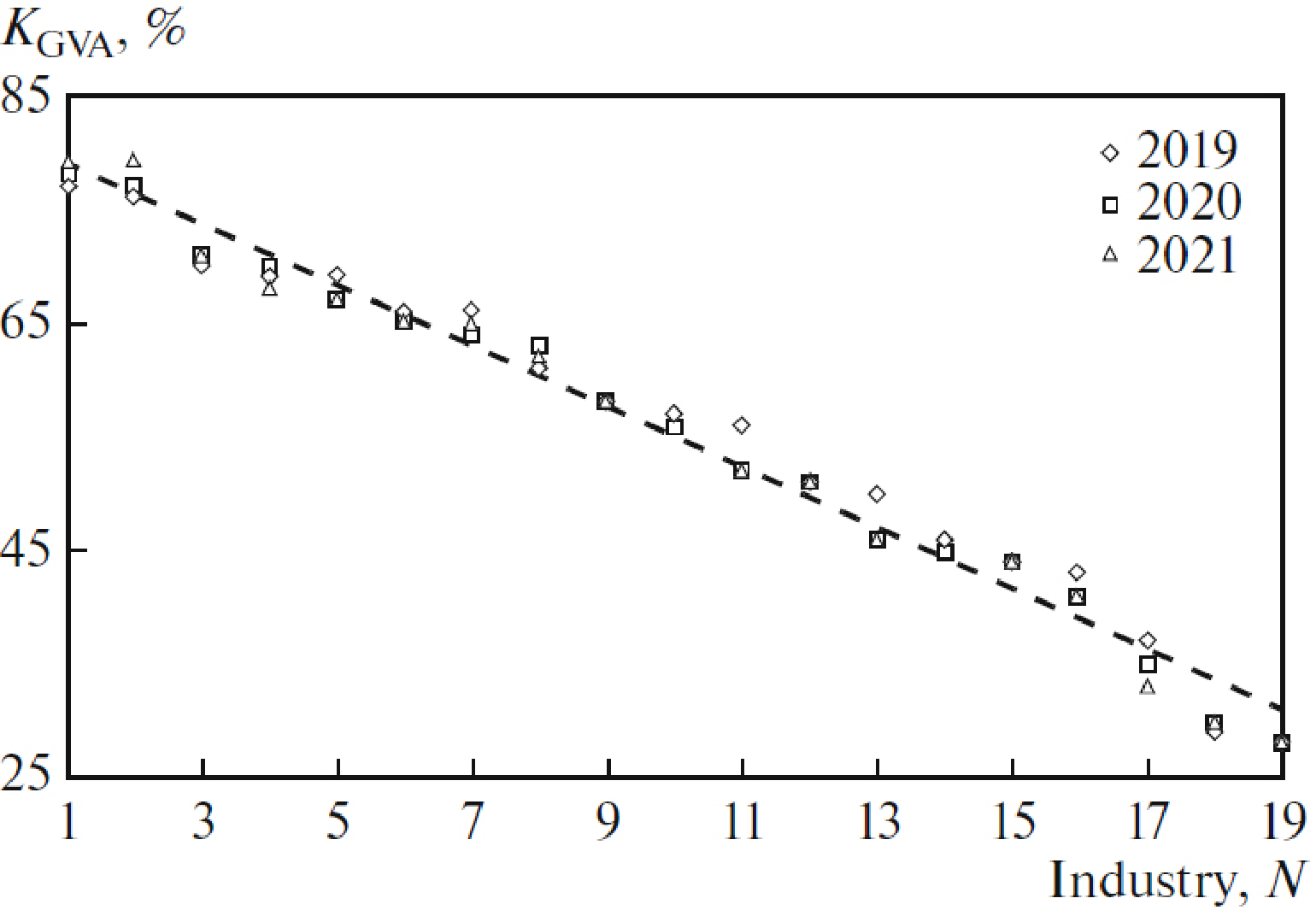}
\caption{Changes in the share of gross value added~\cite{11} of project products $K_\textrm{GVA}$ (equation~\ref{eq1}) by industry ($N$) (Table~\ref{tabl:1}) in comparison with regression values (dotted line) for the period 2019–2021, as of 2022.}
\label{fig1}
\end{figure}

Let us consider changes in relative indicators characterizing the regulatory function of management of high-tech industries. Fig.~\ref{fig1} shows changes in the share of gross value added of project products by industry.

Analysis of the statistical data shown in Fig.~\ref{fig1} has shown that the share of gross value added in basic prices~\cite{11} ranges from 25 to 85\% and changes in a uniform way from industry to industry. This fact indicates a well-balanced competitive environment and effective mechanisms for regulating the profit margin of organizations on the part of the state.

Let us consider changes in absolute indicators characterizing the investment attractiveness of high-tech sectors of the economy using the scree plot as one of the methods of processing economic information in the Data-Mining system in order to identify significant covariates.

\begin{figure}[h]
\centering
\includegraphics[width=\linewidth]{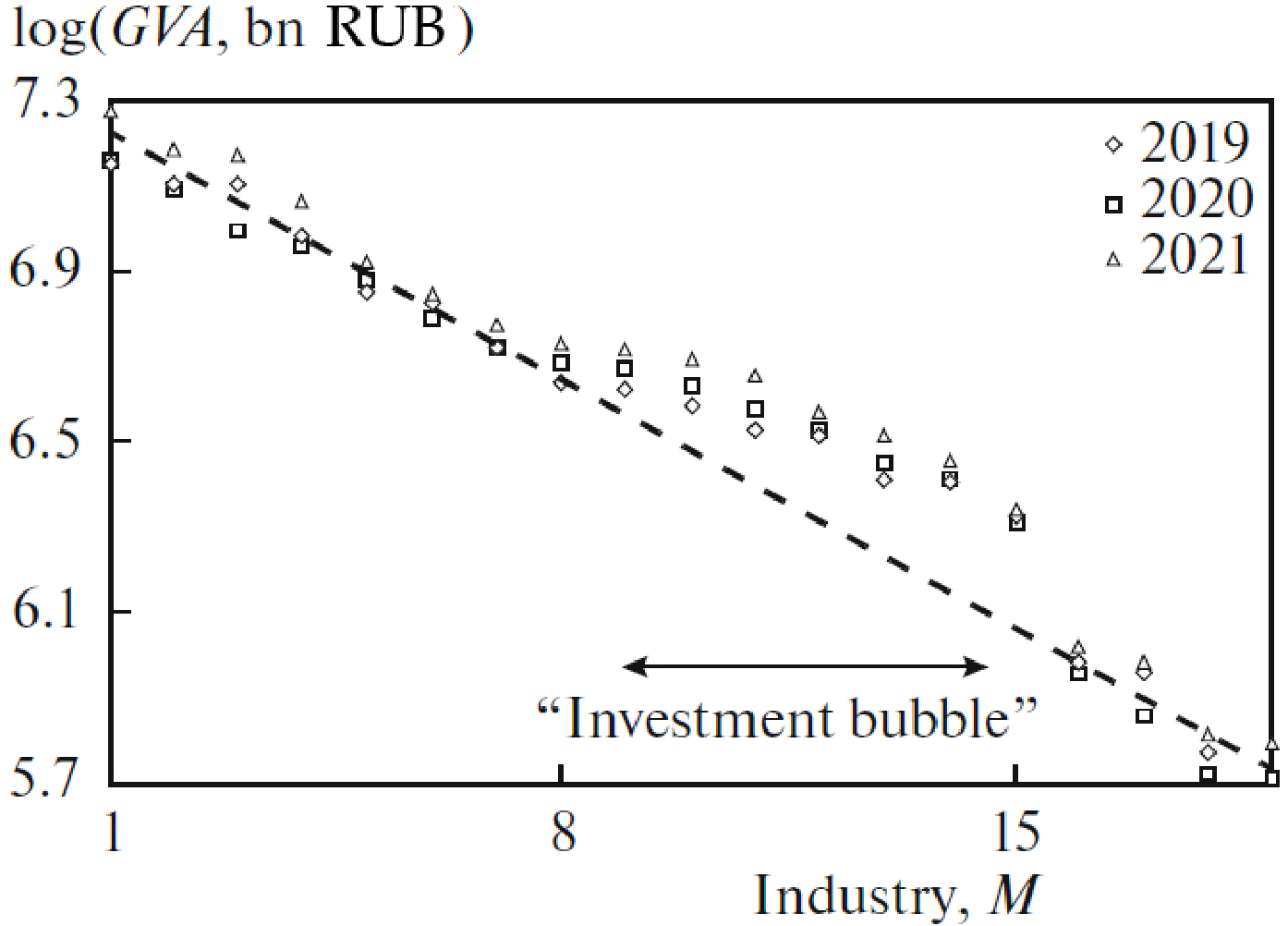}
\caption{Changes in gross value added~\cite{11} by industry ($M$) (Table~\ref{tabl:2}) in comparison with regression values (dotted line) for the period 2019–2021, as of 2022.}
\label{fig2}
\end{figure}

Fig.~\ref{fig2} analyzes changes in gross value added by industry with the aim to investigate possible irregularities in the $GVA$ spectrum.

It was found that the industries listed in Table~\ref{tabl:2} demonstrate a deviation of \textit{GVA} values from the equilibrium trend. An analysis of investment activity has shown that a number of industries create non-equilibrium gross value added, which leads to the formation of an "investment bubble".

\renewcommand\thetable{\arabic{table}}
\begin{table*}[htbp]
\caption{}
  \centering
    \begin{tabular}[t]{|c|l|c|c|}
\hline    
$M$&\multicolumn{1}{c|}{Industry names~\cite{11}}&$K_\textrm{irr}$\\
\hline
1&Professional, scientific and technical activities&0,15\\
\hline
2&Agriculture, forestry and fishing&0,21\\
\hline
3&Human health and social work activities&0,20\\
\hline
4&Education&0,25\\
\hline
5&Information and communication&0,28\\
\hline
6&Electricity, gas, steam and air conditioning
supply&0,31\\
\hline
7&Administrative and support service activities&0,28\\
\hline
    \end{tabular}
\label{tabl:2}    
\end{table*}

The industries that create increased gross value added are analyzed in Fig.~\ref{fig2} and shown in Table~\ref{tabl:2} (sorted by \textit{GVA}), which also includes the values of irregularity coefficient $K_\textrm{irr}$ for \textit{GVA}, determined by the formula~\ref{eq2}:

\begin{equation}
K_\textrm{irr}=log(GVA_\textrm{exp}/GVA_\textrm{reg})
\label{eq2}
\end{equation}

where $GVA_\textrm{exp}$ are the values of gross value added for a given moment in time and $GVA_\textrm{reg}$ is the regression value of gross value added corresponding to an equilibrium economic situation. $K_\textrm{irr}$ shows the scale of "overheating" of the investment sector of the economy in this industry. The analysis has shown that for most industries the value of $K_\textrm{irr}\cong0$, which indicates effective state control of high-tech industries.

\begin{figure*}[h]
\centering

\begin{subfigure}[t]{0.46\linewidth}
\includegraphics[width=8.5cm]{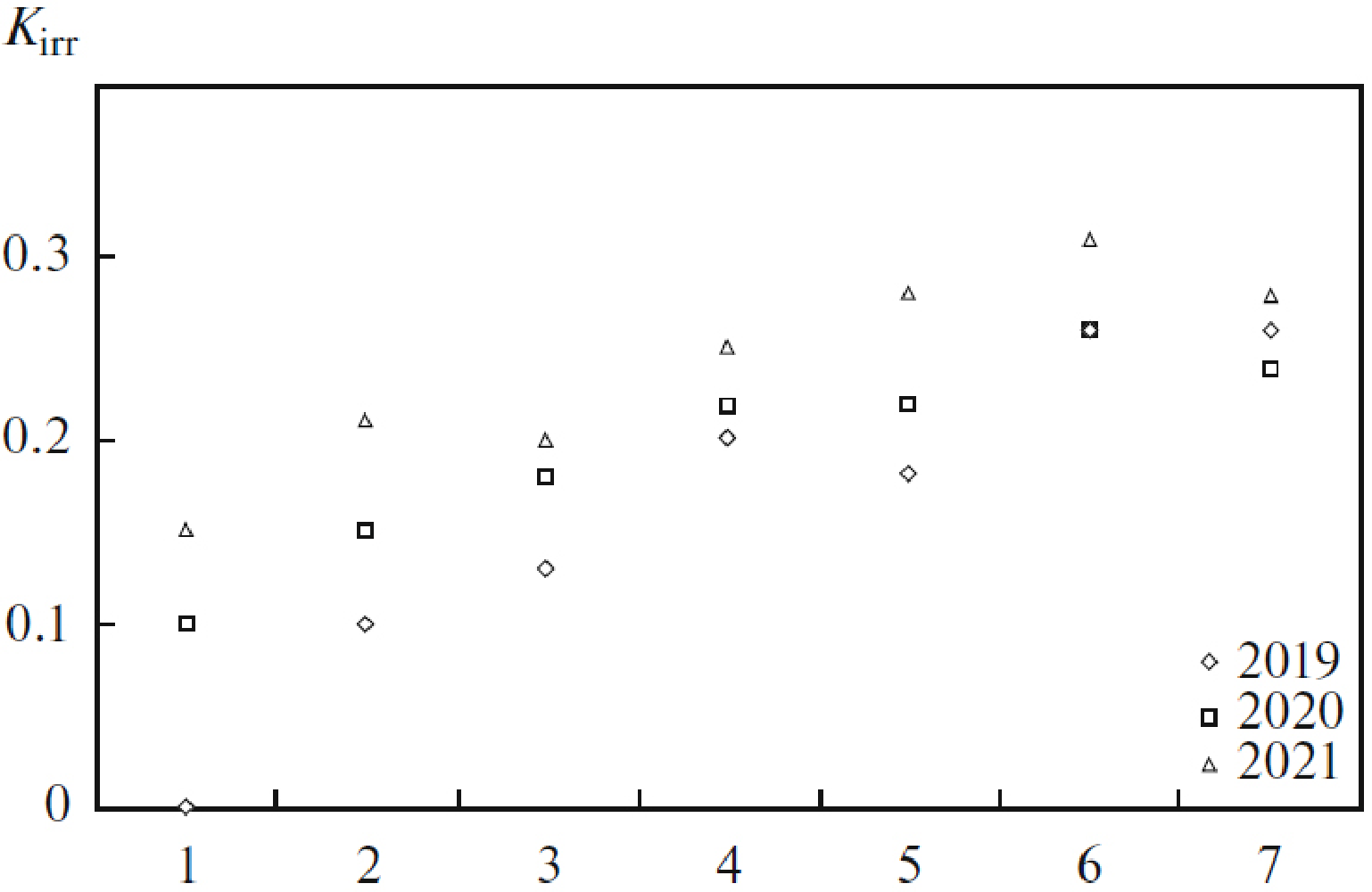}
\centering
\caption{}
\end{subfigure}
\hspace{.5cm}
\begin{subfigure}[t]{0.46\linewidth}
\includegraphics[width=8.5cm]{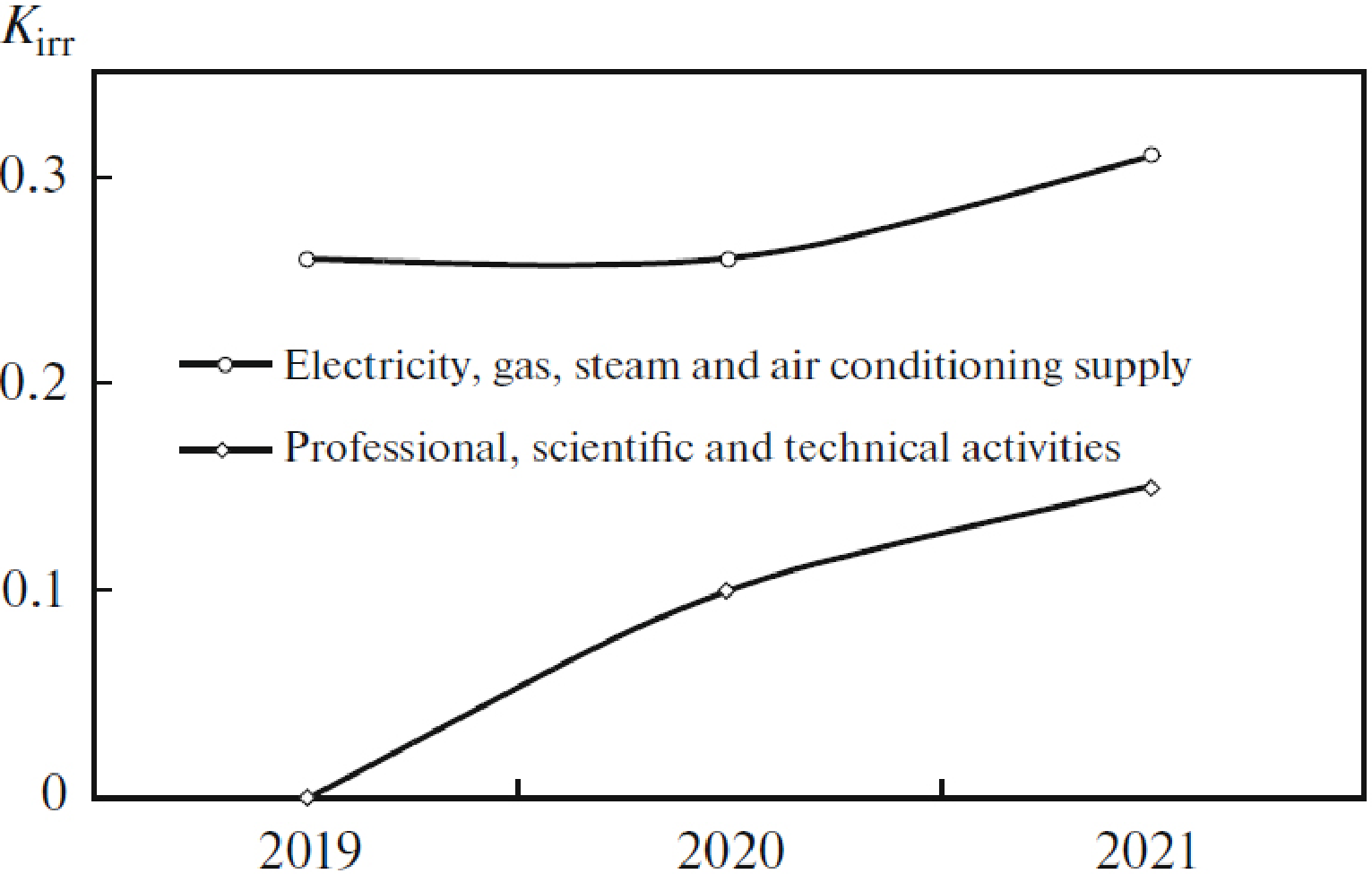}
\centering
\caption{}
\end{subfigure}
\caption{Changes in the coefficient of irregularity of gross value added~\cite{11} by industry~\cite{1,2,3,4,5,6,7} (see Table~\ref{tabl:2})~(a) and for
the industries with the minimum and maximum growth values~(b) for the period 2019–2021, as of 2022.}
\label{fig3}
\end{figure*}

Fig.~\ref{fig3}a shows changes in the $K_\textrm{irr}$ value by industry. Analysis of Fig.~\ref{fig3}a shows that all industries included in the "investment bubble" demonstrate an increase in the value of $K_\textrm{irr}$, which should in the future be corrected by the market, returning the asset to a balanced state. In particular, one of the targets of managing a high-tech enterprise is to prevent "avalanches" of share sales and distortions of information about the real value of assets of enterprises of high-tech industries.

Changes in coefficient $K_\textrm{irr}$ over time for different types of production are analyzed in Fig.~\ref{fig3}b with the aim to identify boundary and forecast values of indicators of the development of high-tech industries. The maximum contribution to economic development has been made by industries engaged in gas infrastructure development, while the minimum contribution corresponds to R\&D-intensive industries. Achieving a balanced economic situation requires development of other high-tech industries, such as manufacturing, manufacture of food products, and others (see Table~\ref{tabl:1}).

{\bf 4.~Results and discussion}\newline
The conducted research has shown that the circumstances regarding high-tech sectors of the economy in the context of import substitution remain problematic. The investment-overheated industries identified as a result of the analysis demonstrate that the Russian economy is in need of diversification. That relates first and foremost to development of industries that manufacture pharmaceuticals, medicinal products, food products, and high technologies.

\end{document}